\documentclass[11pt,a4paper]{article}

\usepackage[T1]{fontenc}
\usepackage[utf8]{inputenc}
\usepackage{lmodern}
\usepackage[margin=3cm]{geometry}
\usepackage{microtype}
\usepackage{amsmath,amssymb}

\usepackage[authoryear,round]{natbib}
\bibpunct{(}{)}{;}{a}{ }{,}

\usepackage[hidelinks]{hyperref}

\title{\textbf{Notes on Crowther and the ``Interpretation'' of Quantum Mechanics (arXiv:2512.14315)}}
\author{Miko\l aj Sienicki\thanks{Polish-Japanese Academy of Information Technology, ul.\ Koszykowa 86, 02-008 Warsaw, Poland, European Union.} \and
Krzysztof Sienicki\thanks{Chair of Theoretical Physics of Naturally Intelligent Systems (NIS), Lipowa 2/Topolowa 19, 05-807 Podkowa Le\'sna, Poland, European Union.}}
\date{December 18, 2025}

\begin{document}
\maketitle

\begin{abstract}
We read Karen Crowther’s \emph{Another 100 Years of Quantum Interpretation?} with two practical goals. First, we spell out what she means by ``interpretation'': an attempt to provide understanding (not just predictions), which may be representationalist or non-representationalist, and which she contrasts with deeper \emph{reductive} (inter-theoretic) explanation---especially in the quantum-gravity setting. Second, we list twelve points where the paper’s physics-facing wording could be sharpened. In our view, several claims are directionally well-motivated but stated more strongly than the underlying physics supports, or they run together distinct notions (e.g.\ ``degrees of freedom,'' ``singularity,'' and different senses of ``locality'') that need careful separation. We end by suggesting that the philosophical question is genuinely worthwhile, but the physics should be phrased more cautiously so that heuristic motivation is not mistaken for strict implication.
\end{abstract}


\section{Introduction: what Crowther means by ``interpretation''}

Crowther treats \emph{interpretations of quantum mechanics} as doing at least two jobs:
(i) tackling problems commonly thought to attach to the theory (above all, the measurement problem), and
(ii) offering understanding of why the theory works and why it has the structure it does \citep[1]{Crowther2025}.

A central move in the paper is to widen the space around ``interpretation'' by stressing that explanation need not be only interpretative. In particular, Crowther contrasts interpretative explanation with \emph{reductive explanation}---explaining a theory’s success, structure, and limitations by embedding it in a newer, more accurate, and/or more fundamental theory \citep[1--3]{Crowther2025}. In the quantum-gravity setting, this leads to a sharp meta-question: if quantum gravity is meant to be deeper than both quantum mechanics and general relativity, why is it often taken for granted that quantum gravity should reductively explain GR, but not quantum mechanics (leaving QM’s ``explanation'' mainly to interpretation) \citep[1--3]{Crowther2025}?

Crowther also uses the now-familiar philosophical contrast between \emph{representationalist} and \emph{non-representationalist} interpretations. Even a non-representationalist stance, she argues, can still deliver understanding (for example, by clarifying inferential roles, licensing practices, or what kind of object a quantum state is supposed to be). On this view, ``interpretation'' is not automatically the same thing as ontology-building \citep[2]{Crowther2025}. She cites pragmatist and related approaches in this direction \citep{Healey2012,Healey2020,Wallace2020}.

Finally, Crowther draws a terminological line that matters for how she counts options: strictly speaking, an ``interpretation'' should not \emph{modify} the standard theory or add new theoretical structure. Views that do add structure (many hidden-variable and collapse proposals) are therefore treated as ``approaches'' or ``alternatives,'' not interpretations \citep[3]{Crowther2025}. Whether one accepts this policing of the term or not, it plays a real role in the paper’s argument.

\section{What do \textit{we} think about ``interpretation''?}
\label{sec:what-do-we-think-interpretation}

In everyday physics talk, ``interpretation'' often does several jobs at once. It helps to separate them explicitly:
\begin{enumerate}
\item \emph{Semantic/representational work}: what (if anything) quantum states, observables, and probabilities represent.
\item \emph{Dynamical/completion work}: whether the standard dynamics is enough for a single-story account of outcomes, or must be supplemented (hidden variables, collapse).
\item \emph{Methodological/heuristic work}: how a conceptual picture guides modelling, unification, or research strategy (for example, toward QFT or quantum gravity).
\end{enumerate}

Crowther’s emphasis sits naturally with the third role: interpretation is not the only route to understanding, because some of what we want from ``interpretation'' might instead be delivered by \emph{reductive explanation}, i.e.\ by embedding QM in a deeper theory \citep{Crowther2025}.

To get a feel for how physicists actually use the term, it is also instructive (with appropriate caution) to look at interpretational polls. Polls do not decide foundational questions; selection effects are enormous. Still, they can be revealing: they show which senses of ``interpretation'' are alive in different communities, and they make it hard to pretend that a single consensus meaning is already in place.

\subsection{Tegmark (1998): an early ``workshop vote'' and the pragmatic mood}
Tegmark reports an explicitly informal poll taken at the 1997 UMBC quantum mechanics workshop \citep{Tegmark1998}. The counts he records are: Copenhagen (13), Many Worlds (8), Bohm (4), Consistent Histories (4), modified dynamics (1), and none/undecided (18) \citep{Tegmark1998}. The point is not to treat these numbers as representative. Rather, they show a familiar attitude: for many working physicists, the predictive machinery is shared, and interpretational preference can look like an extra layer on top of the same calculational core.

Tegmark also emphasizes decoherence as one reason the pressure for literal collapse can feel less urgent (at least in explaining why macroscopic superpositions are not observed). On this reading, a large part of the disagreement is about what counts as an acceptable explanatory story, not about how to compute laboratory statistics.

\subsection{Schlosshauer--Kofler--Zeilinger (2013): the canonical ``snapshot''}
Schlosshauer, Kofler, and Zeilinger’s ``snapshot'' poll (33 participants; 16 questions) became a reference point for later replications and comparisons \citep{SKZ2013}. Its value is not only that it asks for a ``favorite interpretation,'' but also that it maps correlations among attitudes about randomness, measurement, realism, and related themes. Even when one treats it as sociological data rather than scientific evidence, it provides a baseline for discussing how different packages of commitments tend to travel together.

\subsection{Norsen \& Nelson (2013): replication and subcommunity effects}
Norsen and Nelson administered essentially the same questionnaire as \citet{SKZ2013} to attendees at \emph{Quantum Theory Without Observers III} (Bielefeld), obtaining 76 responses \citep{NorsenNelson2013}. Their headline result makes the selection issue unavoidable: in this sample, de Broglie--Bohm gets 63\% of endorsements, while Copenhagen gets 4\%, information-based 5\%, objective collapse 16\%, and ``no preferred interpretation'' 11\% \citep{NorsenNelson2013}. Their broader message is simple: these polls are snapshots of who is in the room.

This paper also highlights an ambiguity that matters for Crowther’s terminology. Many physicists use ``interpretation'' in a broad, theory-adjacent sense (completion strategies and dynamical alternatives are grouped together), even if one prefers to reserve ``interpretation'' for views that do not modify the standard formalism \citep{Crowther2025,NorsenNelson2013}.

\subsection{Sivasundaram \& Nielsen (2016): a broader survey and the knowledge/importance gap}
Sivasundaram and Nielsen surveyed 1234 physicists across eight universities and received 149 responses \citep{SivasundaramNielsen2016}. Their most striking finding is a tension: many respondents report limited familiarity with standard foundational issues, while still saying that interpretations of physical theories matter for understanding \citep{SivasundaramNielsen2016}.

On preferred interpretation, Copenhagen is the modal choice (39\%), while a very large fraction reports no preferred interpretation (36\%); Everett and information-based approaches are each at 6\%, and de Broglie--Bohm is 2\% \citep{SivasundaramNielsen2016}. Outside specialist foundations circles, interpretational commitment is often weaker, but the desire for a coherent conceptual story does not disappear.

\subsection{Jedli\v{c}ka et al.\ (2025): stability, sampling bias, and structure among answers}
Jedli\v{c}ka and coauthors revisit the earlier polls near the centennial of modern quantum mechanics and add a new dataset (40 valid questionnaires overall), together with analyses of dependencies among answers \citep{JedlickaEtAl2025}. Their central descriptive message is stable disagreement rather than convergence. For their preferred-interpretation question (with $N=30$), they report: Copenhagen 60\%, Everett 7\%, de Broglie--Bohm 23\%, and Other 10\% \citep{JedlickaEtAl2025}. They argue that dramatic variations across earlier polls are best explained by sampling and audience effects rather than a clean time trend \citep{JedlickaEtAl2025}.

\subsection{Synthesis: what the polling literature suggests}
Without treating any poll as authoritative, a few lessons are hard to miss:
\begin{enumerate}
\item Pluralism is durable: the surveys do not show convergence.
\item Context matters a lot: different audiences yield very different ``majorities'' \citep{NorsenNelson2013,SivasundaramNielsen2016,JedlickaEtAl2025}.
\item The label ``interpretation'' is used broadly: many physicists include Bohm and collapse proposals under that heading regardless of stricter philosophical taxonomy \citep{Crowther2025,NorsenNelson2013}.
\item Interpretations are valued for understanding even when foundations literacy is limited, at least in broad departmental samples \citep{SivasundaramNielsen2016}.
\end{enumerate}

These facts make Crowther’s methodological suggestion feel natural: if interpretational debate is both important (as a source of understanding) and structurally hard to settle (due to underdetermination and sociological fragmentation), then it is sensible to broaden the explanatory menu, including reductive explanation via deeper theory \citep{Crowther2025}.

\section{Twelve points: potential scientific or conceptual issues in Crowther's paper}
\label{sec:twelve-points}

In what follows we do not dispute Crowther’s overall philosophical question. Our concern is narrower: several physics-facing formulations are likely to be read more strongly than the underlying physics warrants. In each point below, the fix we have in mind is mainly one of \emph{qualification} and \emph{cleaner separation of distinct claims}.

\begin{enumerate}

\item \emph{QM vs QFT and ``many degrees of freedom.''}
Crowther writes that ``standard quantum mechanics is unable to deal with large numbers of degrees of freedom,'' connecting this to fluids and to fields with infinitely many degrees of freedom \citep[11]{Crowther2025}. A gentler and (we think) more accurate phrasing is: nonrelativistic QM can handle arbitrarily many degrees of freedom in principle (i.e.\ any \emph{finite} $N$), but it becomes computationally intractable in practice; \emph{relativistic} QFT, meanwhile, adds further structural commitments (especially Lorentz covariance together with locality/microcausality in the relativistic sense, field excitations, and a natural account of creation/annihilation processes). It is also worth flagging that physicists sometimes use ``QFT'' more loosely for nonrelativistic many-body field formalisms; in that broader usage, ``QFT'' need not be tied to Lorentz invariance, whereas the microcausality/locality constraints are distinctive of \emph{relativistic} QFT. Finally, genuinely \emph{infinite}-degree-of-freedom limits (as in continuum fields and thermodynamic limits) raise additional structural issues (e.g.\ representation-theoretic subtleties and unitarily inequivalent representations) that are not merely matters of ``large $N$.'' As written, Crowther’s sentence risks blending these contrasts.

\item \emph{Variable particle number as ``implied'' by QM + SR.}
Crowther suggests variable particle number is implied by combining quantum mechanics with special relativity and is necessary for scattering with creation/annihilation \citep[11]{Crowther2025}. The motivation is sound in interacting relativistic physics, but the implication language is stronger than needed. A safer statement is that relativistic locality and interaction make a field-theoretic framework natural and (in mainstream relativistic modelling) the \emph{standard} framework for realistic interacting theories, rather than that ``QM + SR'' \emph{logically forces} particle creation in all contexts.

\item \emph{Lorentz invariance and the Schr\"odinger equation.}
Crowther describes Lorentz invariance as ``in conflict'' with the Schr\"odinger equation \citep[11]{Crowther2025}. Nonrelativistic QM is indeed not Lorentz invariant, but this is usually presented as a regime/approximation issue rather than an internal inconsistency. It would read more cleanly to say that the Schr\"odinger dynamics is not the right symmetry framework for relativistic phenomena, which is one reason QFT is required.

\item \emph{Singularity theorems: what they establish.}
Crowther appeals to the Penrose--Hawking singularity theorems to motivate the idea that GR points beyond itself \citep[9]{Crowther2025}. This is standard, but it helps to keep two technical clarifications in view: the theorems typically establish geodesic incompleteness (not automatically curvature blow-up), and the conclusions depend on energy conditions and global assumptions \citep{Penrose1965,HawkingEllis1973}. A small qualification here prevents over-reading.

\item \emph{``Every dynamical field must be quantized'' as a premise.}
Crowther presents the familiar heuristic: quantum theory says every dynamical field must be quantized; GR says spacetime is dynamical; therefore spacetime must be quantized \citep[23]{Crowther2025}. This is a powerful motivation, but the first clause is not a theorem of quantum theory; it is an extrapolation supported by successful unifications plus consistency worries about semiclassical hybrids. It helps to label it explicitly as methodological rather than deductive.

\item \emph{Entropy bounds and a purported ``conflict'' with QFT.}
Crowther notes that black-hole thermodynamics suggests entropy bounds and says this ``conflicts with the predictions of QFT,'' since continuum QFT has infinitely many degrees of freedom per region \citep[12]{Crowther2025}. The underlying \emph{tension} is real and important, but ``conflict'' can sound stronger than intended. Continuum QFT ultraviolet divergences are indeed handled via renormalization, cutoffs, and operational coarse-graining; however, in the specific ``entropy in a bounded region'' discussion, the sharper issue is that the relevant entropies are typically \emph{UV-sensitive} (e.g.\ entanglement entropy is regulator-dependent, and in algebraic QFT the local algebras are of type~III so a naive von Neumann entropy for a sharp region is not straightforwardly defined). The genuine conceptual pressure-point is the tension between continuum QFT’s short-distance structure (arbitrarily many modes at arbitrarily small scales) and the suggestion from gravitational thermodynamics/holography that gravity enforces a finite information capacity for bounded regions \citep{Haag1996,Bousso2002}. To avoid overstatement: this is best read as a structural motivation and an active research programme, not as a settled derivation that straightforwardly ``refutes'' continuum QFT.

\item \emph{AQFT as ``QM + SR without singularities.''}
Crowther characterizes algebraic QFT as a combination of QM and SR ``without any singularities'' \citep[13]{Crowther2025}. AQFT is indeed a mathematically controlled framework that emphasizes locality and structural properties, but it does not simply remove every divergence or pathology one might worry about in QFT. It may be clearer to say that AQFT avoids some perturbative and representation-dependent pitfalls by reformulating the theory in terms of local algebras \citep{Haag1996}.

\item \emph{RG/EFT and the discovery of the Standard Model.}
Crowther suggests renormalization group methods led to EFT and ultimately to the discovery of the Standard Model \citep[11--12]{Crowther2025}. The RG/EFT viewpoint was crucial for understanding and legitimizing renormalized QFTs, but the Standard Model also emerged through symmetry principles, renormalizability constraints, and experiment-driven model building in a more intertwined way. The point is directionally right, but the causal wording can be softened \citep{Wilson1971}.

\item \emph{Everett as ``local and deterministic.''}
Deterministic is standard for Everett (unitary evolution), but ``local'' is slippery \citep[19]{Crowther2025}. It is safer to separate at least three notions that are often conflated: (i) \emph{no superluminal signaling} (standard in quantum theory), (ii) \emph{Bell locality/factorizability} (violated by quantum correlations), and (iii) \emph{local micro-dynamics} in relativistic QFT (often formalized via microcausality of observables at spacelike separation). A small qualification along these lines would prevent readers from thinking there is a single settled locality claim on the table \citep{Wallace2012}.

\item \emph{Gravity-mediated entanglement proposals and Everett.}
Crowther notes that some gravity-mediated entanglement proposals have been framed assuming Everett \citep[20]{Crowther2025}. Even when particular authors motivate these experiments via Everett, the core logic is often presented as broadly interpretation-neutral but \emph{conditional}: under standard experimental and modelling assumptions (e.g.\ no hidden classical communication channel between the laboratories, no post-selection loophole that effectively conditions on rare events, and an interaction structure that is local in the relevant operational sense), it is often argued that entanglement generation between two systems via a mediator rules out treating the mediator as ``purely classical'' (roughly: as mere shared classical randomness or classical information passed by local operations). This is the motivation behind the well-known ``gravity-mediated entanglement'' proposals \citep{BoseEtAl2017,MarlettoVedral2017}. However, the precise scope of the ``witness'' claim and the background assumptions under which it goes through have been actively discussed (including models that aim to reproduce entanglement generation with a classical gravitational sector, and analyses that qualify what the observation would and would not establish) \citep{ReginattoHall2018,MartinMartinezEtAl2023,HuggettLinnemannSchneider2023}. It therefore helps to distinguish (i) motivational packaging (including Everett-friendly narratives) from (ii) the conditional operational inference, and to flag explicitly that the inference is assumption-sensitive and not universally taken to be ``automatic.''

\item \emph{Many-worlds and the inflationary multiverse.}
Crowther reports the Bousso--Susskind idea that the many worlds of QM and the many worlds of eternal inflation are ``the same'' \citep[20]{Crowther2025}. This is an interesting unificatory speculation, but it does not function as a standard ``solution'' to the measurement problem; it shifts interpretive burdens into cosmological postulates and measures \citep{BoussoSusskind2012}.

\item \emph{String theory as a reductive explanation of QFT.}
Crowther writes that string theory provides a more fundamental framework than QFT and reductively explains QFT and particular QFTs \citep[21]{Crowther2025}. This is a reasonable research-program narrative (many QFT structures appear as limits or sectors of string theory), but it is safer to state it in a \emph{case-by-case} way: in many examples, string theory yields controlled constructions and dualities whose low-energy/long-distance limits reproduce particular QFTs, and in that sense can provide reductive explanations for those sectors. Still, the scope of this ``reductive explanation'' claim is not uniform across all QFT practice, and it remains conditional on the (still unsettled) status of string theory as the correct description of nature.
\end{enumerate}

\section{Feyerabend vs.\ Popper/Kuhn/Lakatos on Quantum Mechanics}
\label{sec:feyerabend-popper-kuhn-lakatos-qm}

Philosophers of science disagree not only about which interpretation of quantum mechanics is best, but also about what counts as a good scientific question in the first place. This matters because many interpretational debates are shaped by empirical equivalence (at least in their standard forms), so arguments often lean on explanatory, methodological, or historical considerations. Popper, Kuhn, Lakatos, and Feyerabend offer four useful lenses for thinking about why interpretation persists and what progress would look like \citep{SEP_Popper,SEP_Kuhn,SEP_Lakatos,SEP_Feyerabend}.

\subsection{Popper: falsifiability, realism, and the push toward risky tests}
Popper’s core norm is methodological: a proposal is scientific to the extent that it exposes itself to potential refutation \citep{Popper1959,SEP_Popper}. In the QM context, this tends to generate a familiar Popperian worry: if two interpretations never differ on any possible observation, then the dispute risks becoming extra-empirical.

Popper would not say such work is useless. He allows that metaphysical research programmes can be heuristically valuable. But his preferred endpoint is still a conjecture that leads to risky tests or at least to a clearer separation between what is empirical content and what is added commentary \citep{Popper1962,SEP_Popper}.

Applied to the measurement problem, this attitude keeps pressure on foundations work to become crisp physics: either a completion strategy with new commitments, or a dynamical modification with observable consequences, rather than a purely verbal dissolution. Popper is also suspicious of methodological moves that declare whole classes of questions meaningless, since that can function as insulation against criticism \citep{SEP_Popper}.

\subsection{Kuhn: paradigms, normal science, and why pluralism lasts}
Kuhn’s account is historical: science alternates between normal science (puzzle-solving within a shared paradigm) and revolutionary episodes (paradigm change) \citep{Kuhn1970,SEP_Kuhn}. On a Kuhnian reading, textbook QM and especially QFT practice look like a stable and extraordinarily productive paradigm. That helps explain why interpretational disagreement can persist: normal science can proceed without a single agreed ontology, so long as the exemplar-based problem-solving machinery works.

In this picture, the measurement problem can remain a standing anomaly without triggering a crisis, because it does not typically disrupt the day-to-day successes of the paradigm. Kuhn would expect large changes to be driven by shifts in the broader theoretical setting (for instance, quantum gravity or cosmology) where the standard ``measurement'' talk becomes hard to formulate cleanly \citep{Kuhn1970,SEP_Kuhn}.

\subsection{Lakatos: research programmes and progressive problemshifts}
Lakatos offers a middle position. Science advances via competing research programmes with a protected hard core and a flexible protective belt. What matters is whether a programme is progressive (it leads to new predictions or fruitful problemshifts) or degenerating (it survives mainly by patching) \citep{Lakatos1970,SEP_Lakatos}.

In QM foundations, it is natural to treat leading interpretations as programmes: Everett-type views protect unitarity; Bohm-type views protect definite configurations; collapse-type views protect definite outcomes via modified dynamics; operationalist variants protect an instrumental core plus a measurement cut. A key Lakatosian point is that it can be rational to work within a programme even before decisive experimental separation, provided it is theoretically progressive: it unifies domains, clarifies puzzles, yields new models, or points toward new tests \citep{Lakatos1970,SEP_Lakatos}. This sits comfortably with Crowther’s emphasis on heuristic value.

\subsection{Feyerabend: pluralism as a safeguard against premature closure}
Feyerabend rejects the idea of a single universal scientific method and defends methodological pluralism---famously summarized as ``anything goes''---as a safeguard against dogmatism and premature closure \citep{Feyerabend1975,SEP_Feyerabend}. Read charitably, the slogan is not a celebration of chaos; it is a reminder that historically, major advances often required violating reigning methodological rules.

In QM foundations, Feyerabend’s stance translates into a strong defence of interpretational diversity. Rival pictures force us to expose what is assumed rather than observed, and they help prevent an orthodoxy from declaring questions illegitimate simply because they are difficult or unfashionable.

\subsection{What these philosophies predict about QM interpretation debates}
These four viewpoints suggest different ``equilibria'' for quantum foundations:
\begin{itemize}
\item Popper predicts continued pressure toward testable differences or explicit acknowledgement that parts of the debate are (for now) extra-empirical \citep{Popper1959,SEP_Popper}.
\item Kuhn predicts durable pluralism under a successful paradigm, with major change more likely after a context shift or crisis \citep{Kuhn1970,SEP_Kuhn}.
\item Lakatos predicts rational coexistence of programmes, evaluated by progressive vs.\ degenerating problemshifts \citep{Lakatos1970,SEP_Lakatos}.
\item Feyerabend predicts (and endorses) proliferation, treating heterodoxy as an epistemic resource \citep{Feyerabend1975,SEP_Feyerabend}.
\end{itemize}

\subsection{A practical upshot for Crowther-style ``interpretation''}
Crowther’s proposal---to treat interpretation partly as a heuristic tool and to take reductive explanation seriously in quantum gravity---fits naturally with a Lakatosian way of talking and is compatible with Kuhn’s emphasis on context shifts. A Popperian will ask for riskier, more test-connected payoffs; a Feyerabendian will value pluralism even when the payoffs are not immediately test-discriminating.

Related reflections on methodological pluralism in the context of AI-assisted scientific work are discussed in \citet{SienickiSienicki2025AIAgainstMethodZenodo}.

\section{Summary}

Crowther’s paper raises a worthwhile meta-question about why reductive explanation is so often demanded for GR but not for QM in the quantum-gravity programme \citep{Crowther2025}. Our main recommendation is modest: several physics-facing formulations would benefit from slightly more careful qualification, so that strict implication, plausible heuristic motivation, and speculative unification proposals are kept clearly distinct.

% ------------------ Bibliography (manual \bibitem, Chicago-like formatting) ------------------


\begin{thebibliography}{99}

\bibitem[Bose et~al.(2017)]{BoseEtAl2017}
Bose, Sougato, Anupam Mazumdar, Gavin W. Morley, Hendrik Ulbricht, Mauro Toro\v{s}, Michael Paternostro, Andrew A. Geraci, Peter F. Barker, M. S. Kim, and Gerard Milburn. 2017.
``Spin Entanglement Witness for Quantum Gravity.''
\emph{Physical Review Letters} 119: 240401.
\href{https://doi.org/10.1103/PhysRevLett.119.240401}{doi:10.1103/PhysRevLett.119.240401}.

\bibitem[Bousso(2002)]{Bousso2002}
Bousso, Raphael. 2002. ``The Holographic Principle.'' \emph{Reviews of Modern Physics} 74(3): 825--874.
\href{https://doi.org/10.1103/RevModPhys.74.825}{doi:10.1103/RevModPhys.74.825}.

\bibitem[Bousso and Susskind(2012)]{BoussoSusskind2012}
Bousso, Raphael, and Leonard Susskind. 2012. ``Multiverse Interpretation of Quantum Mechanics.''
\emph{Physical Review D} 85: 045007.
\href{https://doi.org/10.1103/PhysRevD.85.045007}{doi:10.1103/PhysRevD.85.045007}.

\bibitem[Crowther(2025)]{Crowther2025}
Crowther, Karen. 2025. ``Another 100 Years of Quantum Interpretation?''
\emph{arXiv} preprint arXiv:2512.14315v1 [physics.hist-ph], December 2025.
\href{https://doi.org/10.48550/arXiv.2512.14315}{doi:10.48550/arXiv.2512.14315}.
\newblock \url{https://arxiv.org/abs/2512.14315}.

\bibitem[Feyerabend(1975)]{Feyerabend1975}
Feyerabend, Paul. 1975.
\emph{Against Method: Outline of an Anarchistic Theory of Knowledge}.
London: New Left Books.

\bibitem[Haag(1996)]{Haag1996}
Haag, Rudolf. 1996. \emph{Local Quantum Physics: Fields, Particles, Algebras}. 2nd ed. Berlin: Springer.

\bibitem[Hawking and Ellis(1973)]{HawkingEllis1973}
Hawking, Stephen W., and George F. R. Ellis. 1973. \emph{The Large-Scale Structure of Space-Time}. Cambridge: Cambridge University Press.

\bibitem[Healey(2012)]{Healey2012}
Healey, Richard. 2012. ``Quantum Theory: A Pragmatist Approach.''
\emph{The British Journal for the Philosophy of Science} 63(4): 729--771.
\href{https://doi.org/10.1093/bjps/axr054}{doi:10.1093/bjps/axr054}.

\bibitem[Healey(2020)]{Healey2020}
Healey, Richard. 2020. ``Pragmatist Quantum Realism.''
In \emph{Scientific Realism and the Quantum}, edited by Steven French and Juha Saatsi, 123--146.
Oxford: Oxford University Press.

\bibitem[Huggett, Linnemann, and Schneider(2023)]{HuggettLinnemannSchneider2023}
Huggett, Nick, Niels Linnemann, and Mike D. Schneider. 2023. \emph{Quantum Gravity in a Laboratory?}
Cambridge: Cambridge University Press.
\href{https://doi.org/10.1017/9781009327541}{doi:10.1017/9781009327541}.

\bibitem[Jedli\v{c}ka et~al.(2025)]{JedlickaEtAl2025}
Jedli\v{c}ka, Petr O., \v{S}imon Kos, Martin \v{S}m\'id, Ji\v{r}\'i Vomlel, and Jan Slav\'ik. 2025.
``Has Anything Changed? Tracking Long-Term Interpretational Preferences in Quantum Mechanics.''
\emph{arXiv} preprint arXiv:2507.09988 [quant-ph].
\href{https://doi.org/10.48550/arXiv.2507.09988}{doi:10.48550/arXiv.2507.09988}.
\newblock \url{https://arxiv.org/abs/2507.09988}.

\bibitem[Kuhn(1970)]{Kuhn1970}
Kuhn, Thomas S. 1970. \emph{The Structure of Scientific Revolutions}. 2nd ed., enlarged. Chicago: University of Chicago Press.

\bibitem[Lakatos(1970)]{Lakatos1970}
Lakatos, Imre. 1970. ``Falsification and the Methodology of Scientific Research Programmes.''
In \emph{Criticism and the Growth of Knowledge}, edited by Imre Lakatos and Alan Musgrave.
Cambridge: Cambridge University Press.

\bibitem[Marletto and Vedral(2017)]{MarlettoVedral2017}
Marletto, Chiara, and Vlatko Vedral. 2017.
``Gravitationally Induced Entanglement between Two Massive Particles Is Sufficient Evidence of Quantum Effects in Gravity.''
\emph{Physical Review Letters} 119: 240402.
\href{https://doi.org/10.1103/PhysRevLett.119.240402}{doi:10.1103/PhysRevLett.119.240402}.

\bibitem[Mart\'{\i}n-Mart\'{\i}nez and Perche(2023)]{MartinMartinezEtAl2023}
Mart\'{\i}n-Mart\'{\i}nez, Eduardo, and T.\ Rick Perche. 2023.
``What Gravity Mediated Entanglement Can Really Tell Us about Quantum Gravity.''
\emph{Physical Review D} 108: L101702.
\href{https://doi.org/10.1103/PhysRevD.108.L101702}{doi:10.1103/PhysRevD.108.L101702}.
\newblock \href{https://doi.org/10.48550/arXiv.2208.09489}{arXiv:2208.09489}.

\bibitem[Norsen and Nelson(2013)]{NorsenNelson2013}
Norsen, Travis, and Sarah Nelson. 2013.
``Yet Another Snapshot of Foundational Attitudes Toward Quantum Mechanics.''
\emph{arXiv} preprint arXiv:1306.4646 [quant-ph].
\href{https://doi.org/10.48550/arXiv.1306.4646}{doi:10.48550/arXiv.1306.4646}.
\newblock \url{https://arxiv.org/abs/1306.4646}.

\bibitem[Penrose(1965)]{Penrose1965}
Penrose, Roger. 1965. ``Gravitational Collapse and Space-Time Singularities.''
\emph{Physical Review Letters} 14: 57--59.
\href{https://doi.org/10.1103/PhysRevLett.14.57}{doi:10.1103/PhysRevLett.14.57}.

\bibitem[Popper(1959)]{Popper1959}
Popper, Karl R. 1959. \emph{The Logic of Scientific Discovery}. London: Hutchinson.

\bibitem[Popper(1962)]{Popper1962}
Popper, Karl R. 1962. \emph{Conjectures and Refutations: The Growth of Scientific Knowledge}. New York: Basic Books.

\bibitem[Reginatto and Hall(2018)]{ReginattoHall2018}
Reginatto, Marcel, and Michael J.\ W.\ Hall. 2018.
``Entanglement of Quantum Fields via Classical Gravity.''
\emph{arXiv} preprint arXiv:1809.04989 [quant-ph].
\href{https://doi.org/10.48550/arXiv.1809.04989}{doi:10.48550/arXiv.1809.04989}.
\newblock \url{https://arxiv.org/abs/1809.04989}.

\bibitem[Schlosshauer, Kofler, and Zeilinger(2013)]{SKZ2013}
Schlosshauer, Maximilian, Johannes Kofler, and Anton Zeilinger. 2013.
``A Snapshot of Foundational Attitudes Toward Quantum Mechanics.''
\emph{Studies in History and Philosophy of Modern Physics} 44(3): 222--230.
\href{https://doi.org/10.1016/j.shpsb.2013.04.004}{doi:10.1016/j.shpsb.2013.04.004}.
\newblock \url{https://arxiv.org/abs/1301.1069}.

\bibitem[Sienicki and Sienicki(2025)]{SienickiSienicki2025AIAgainstMethodZenodo}
Sienicki, Miko\l aj, and Krzysztof Sienicki. 2025.
\newblock ``AI Against Method: An Anarchist Essay.''
\newblock \emph{Zenodo} (Publication), version v1.
\newblock \href{https://doi.org/10.5281/zenodo.17970889}{doi:10.5281/zenodo.17970889}.
\newblock \url{https://zenodo.org/records/17970889}.

\bibitem[Sivasundaram and Nielsen(2016)]{SivasundaramNielsen2016}
Sivasundaram, Sujeevan, and Kristian Hvidtfelt Nielsen. 2016.
``Surveying the Attitudes of Physicists Concerning Foundational Issues of Quantum Mechanics.''
\emph{arXiv} preprint arXiv:1612.00676 [physics.hist-ph].
\href{https://doi.org/10.48550/arXiv.1612.00676}{doi:10.48550/arXiv.1612.00676}.
\newblock \url{https://arxiv.org/abs/1612.00676}.

\bibitem[Tegmark(1998)]{Tegmark1998}
Tegmark, Max. 1998. ``The Interpretation of Quantum Mechanics: Many Worlds or Many Words?''
\emph{Fortschritte der Physik} 46(6--8): 855--862.
doi:10.1002/(SICI)1521-3978(199811)46:6/8<855::AID-PROP855>3.0.CO;2-Q.
\newblock \url{https://arxiv.org/abs/quant-ph/9709032}.

\bibitem[Wallace(2012)]{Wallace2012}
Wallace, David. 2012. \emph{The Emergent Multiverse: Quantum Theory According to the Everett Interpretation}. Oxford: Oxford University Press.

\bibitem[Wallace(2020)]{Wallace2020}
Wallace, David. 2020. ``On the Plurality of Quantum Theories: Quantum Theory as a Framework, and Its Implications for the Quantum Measurement Problem.''
In \emph{Scientific Realism and the Quantum}, edited by Steven French and Juha Saatsi, 78--102.
Oxford: Oxford University Press.

\bibitem[Wilson(1971)]{Wilson1971}
Wilson, Kenneth G. 1971. ``Renormalization Group and Critical Phenomena. I. Renormalization Group and the Kadanoff Scaling Picture.''
\emph{Physical Review B} 4: 3174--3183.
\href{https://doi.org/10.1103/PhysRevB.4.3174}{doi:10.1103/PhysRevB.4.3174}.

% ---- Stanford Encyclopedia of Philosophy (use substantive revision year) ----

\bibitem[Oberheim(2016)]{SEP_Feyerabend}
Oberheim, Eric. 2016. ``Paul Feyerabend.'' In \emph{The Stanford Encyclopedia of Philosophy}.
(First published 1997; substantive revision October 28, 2016.)
\newblock \url{https://plato.stanford.edu/entries/feyerabend/}.

\bibitem[Bird(2025)]{SEP_Kuhn}
Bird, Alexander. 2025. ``Thomas Kuhn.'' In \emph{The Stanford Encyclopedia of Philosophy}.
(First published 2004; substantive revision September 2, 2025.)
\newblock \url{https://plato.stanford.edu/entries/thomas-kuhn/}.

\bibitem[Musgrave and Pigden(2021)]{SEP_Lakatos}
Musgrave, Alan, and Charles Pigden. 2021. ``Imre Lakatos.'' In \emph{The Stanford Encyclopedia of Philosophy}.
(First published 2016; substantive revision August 18, 2021.)
\newblock \url{https://plato.stanford.edu/entries/lakatos/}.

\bibitem[Thornton(2022)]{SEP_Popper}
Thornton, Stephen. 2022. ``Karl Popper.'' In \emph{The Stanford Encyclopedia of Philosophy}.
(First published 1997; substantive revision April 22, 2022.)
\newblock \url{https://plato.stanford.edu/entries/popper/}.

\end{thebibliography}
\end{document}